\documentclass[aip,twocolumn,reprint]{revtex4-1}
\usepackage{graphicx}
\begin{document}
\title{Frequency-dependent polarization-angle-phase-shift in the microwave-induced magnetoresistance oscillations}
\author{Han-Chun Liu}
\author{Tianyu Ye}
\affiliation{Department of Physics and Astronomy, Georgia State University, Atlanta, Georgia 30303}
\author{W. Wegscheider}
\affiliation{Laboratorium f\"ur Festk\"orperphysik, ETH Z\"urich, CH-8093 Z\"urich, Switzerland}
\author{R. G. Mani}
\affiliation{Department of Physics and Astronomy, Georgia State University, Atlanta, Georgia 30303}\date{\today}
\begin{abstract}
Linear polarization angle, $\theta$, dependent measurements of the microwave radiation-induced oscillatory magnetoresistance, $R_{xx}$,  in high mobility GaAs/AlGaAs 2D electron devices have shown a $\theta$ dependence in the oscillatory amplitude along with magnetic field, frequency, and extrema-dependent phase shifts, $\theta_{0}$.  Here, we suggest a microwave frequency dependence of  $\theta_{0} (f)$ using an analysis that averages over other smaller contributions, when those contributions are smaller than estimates of the experimental uncertainty.
\end{abstract}
\maketitle
\section{Introduction}
The quasi two-dimensional electron system (2DES) in high mobility GaAs/AlGaAs heterostructures has served to identify a new type of zero-resistance state in the 2DES, one which occurs at high filling factors or low magnetic fields under microwave photo-excitation.  \cite{2002RGManiNature,ZudovPRLDissipationless2003} Such zero-resistance states have been fascinating in part because they could help to point out the necessary conditions for obtaining vanishing resistance in the 2DES in a magnetic field. The associated oscillatory effect is also interesting from the application perspective since it could lead to new approaches for a frequency and power sensitive radiation detector in the microwave and terahertz bands, for a large number of  applications.\cite{2002SiegelPH, 1995HuOpticsExp, 2008RGManiAPL}

Numerous experimental \cite{2002RGManiNature,ZudovPRLDissipationless2003, 2004RGManiPRB,2004RGManiPRL,2004SAStudenikinSSComms, 2004RGManiPhysicaE,2004AEKovalevSSComms, 2004RRDuPhysicaE,2005JHSmetPRL,2007AWirthmannPRB,2007SAStudenikinPRB,2008RGManiAPL,2008OERaichevPRB, 2008SWiedmannPRB,2009RGManiPRB,2009LCTungSSComms,2010RGManiPRB,2010ANRamanayakaPRB,2010SWiedmannPRB,2010OMFedorychPRB,2010SWiedmannPRL,2011ANRamanayakaPRB,
2011RGManiPRB,2012ANRamanayakaPRB,2012ABoganPRB, 2012RGManiNComms,2013SDietrichJAP,2013RGManiPRB,2013RGManiSRep,2013RGManiSRep1,2013TYeAPL,2013ATHatkePRB,2013DKonstantinovPRL,2014TYePRB}and theoretical\cite{2003ACDurstPRL,2003VRyzhiiJphyConMat,2003XLLeiPRL,2004XLLeiJPhysConMat,2004KPrakPRB,2003AAKoulakovPRB,2003AVAndreevPRL,2003VRyzhiiPRB,2004PHRiveraPRB,
2004SAMikhailovPRB,2005AAuerbachPRL,
2005JInarreaPRL,2005XLLeiPRB,2005IADmitrievPRB,2006JInarreaAPL,2006JDietelPRB,2007JInarreaPRB,2007ADChepelianskiiEurPhysJB,2008JInarreaPRB,2008SWangPRB,2008JInarreaAPL, 2009JInarreaAPL,2009IGFinklerPRB,
2009PHRiveraPRB,2009XLLeiAPL,2010JInarreaPRB,2010JInarreaNanotechnology,2011JInarreaPRB,2011JInarreaAPL,2011SAMikhailovPRB,
2012JInarreaAPL,2012ADChepelianskiiPRB,2012XLLeiPRB,2013JInarreaPhysLettA,2013JInarreaJAP,
2013AKunoldPhysB,2013OVZhirovPRB,2014JInarreaEPL,2014XLLeiJAP} works about the radiation-induced magnetoresistance oscillations (RIMO) and associated zero-resistance states have been published over the past decade. At the present, it is understood that RIMOs are “$1/4$-cycle phase-shifted” in $1/B$, and the oscillatory minima occur in $R_{xx}$ vs. $B$ plots broadly about $B=[4/(4j+1)]B_f$, where $B_f=2\pi fm^*/e$, $f$ is the microwave frequency, $m^*$ is the effective electron mass in GaAs and $j=1, 2, 3$\dots. The theoretically proposed physical mechanisms for RIMOs include the displacement model\cite{2003ACDurstPRL,2003XLLeiPRL}, the nonparabolicity model\cite{2003AAKoulakovPRB}, the inelastic model\cite{2005IADmitrievPRB} and the radiation-driven electron-orbit model\cite{2005JInarreaPRL,2007JInarreaPRB}. These different theories have suggested dissimilar behavior of some physical properties  such as, for example the polarization-angle-dependence and the power-dependence of the RIMOs. As a result, an interesting issue concerns the sensitivity of the RIMOs to the polarization angle of linearly polarized microwaves. Early linear microwave polarization sensitivity work carried out on L-shaped samples\cite{2004RGManiPhysicaE} showed that the the frequency and the phase of RIMOs are insensitive to the linear polarization angle of the microwaves. Later work indicated that  RIMOs were insensitive to circularly and linearly polarized microwaves in square-shaped samples in a quasioptical measurement\cite{2005JHSmetPRL}. More recently, a polarization-angle-dependence in the amplitude of RIMOs has been demonstrated\cite{2011RGManiPRB,2012ANRamanayakaPRB,2014TYePRB}. The results were roughly consistent with the predictions of the displacement model, the nonparabolicity model, and the radiation-driven electron-orbit model for $\gamma<\omega$, where $\gamma$ is damping factor and $\omega=2\pi f$.\cite{2003ACDurstPRL,2003XLLeiPRL,2003AAKoulakovPRB,2005JInarreaPRL,2007JInarreaPRB,2012JInarreaAPL} Finally, Ramanayaka \textit{et al.}  showed that the $R_{xx}$ varied sinusoidally vs. $\theta$ at low microwave power, following the empirical relation $R_{xx}(\theta)=A\pm Ccos^2(\theta -\theta_0)$, where $\theta$ is microwave polarization angle, $\theta_0$ is phase shift, and the plus and minus signs corresponded to the oscillatory maxima and minima, respectively.\cite{2012ANRamanayakaPRB} The results suggested both  a $f$-dependence and a $B$-dependence in $\theta_{0}$, although the phase shifts did not appear to be systematically responsive to any experimental parameters.  These studies indicated that the observed phase shifts required further experimental investigation.

Thus, we extract the frequency dependence of the phase shift in the $R_{xx}$ vs. $\theta$ results by applying an analysis that averages over other smaller contributions, when those contributions are smaller than estimates of the experimental uncertainty. The results suggest a non-vanishing frequency dependence in the phase shift, i.e., $\theta_{0} = \theta_{0}(f)$, over the frequency interval $32 \le f \le 50$ GHz.  

\section{experiment and results}
The experimental setup for the polarization-orientation measurements is shown in Fig.\ 1(a). Linearly polarized  microwaves  are generated by an antenna inside a rotatable microwave launcher, and they are transmitted via a cylindrical waveguide to the sample. The samples consist of 400-$\mu m$-wide Hall bar with alloyed gold-germanium  contacts fabricated from  GaAs/AlGaAs heterojunctions with a 2DES with carrier density $\approx2.7\times10^{11}cm^{-2}$ and mobility $\approx 8\times10^6cm^{2}\cdot V^{-1}\cdot s^{-1}$ .  The samples are immersed in pumped liquid helium to maintain a temperature of 1.5 K during the measurements. Standard four-terminal lock-in techniques are utilized to measure the diagonal resistance $R_{xx}$.  Finally, the polarization angle $\theta$ is defined as the angle between the microwave electric field $E$ and the Hall bar axis. Thus, in experiment, the gradual increase of $\theta$ from $0^\circ$ to $360^\circ$ is achieved by rotating  the microwave launcher.

At the outset to the experiment, a power detector is connected to a power meter and placed at the end of cylindrical waveguide, see Fig.\ 1(c) inset. This power detector is sensitive to the radiation field along its preferred axis. The orientation of this detector is fixed parallel to the antenna, setting $\theta=0^\circ$. Then, the antenna is turned from $0^\circ$ to $360^\circ$ at $5^\circ$ increment for a number of different frequencies, $f$, from $32$ GHz to $50$ GHz, to characterize the linear polarization angle of the microwaves at the bottom of the waveguide sample holder. Fig.\ 1(c) shows the normalized detected power as a function of $\theta$ at $40.791$ GHz. As expected for linearly polarized microwaves, the detected power varies sinusoidally with $\theta$, and this sinusoidal curve can be described by $P=A+C\cos^2(\theta -\theta_0)$, where $P$ is the detected power, $A$ is the dark response without microwaves, and $C$ is the amplitude of cosine square function. The cosine square function has been used here because the microwave power is proportional to the square of the microwave electric field. Such data fits to this sinusoidal function were used to extract a phase shift, $\theta_0$, in the absence of a sample to characterize the polarization angle in the measurement setup. Fig.\ 1(c) demonstrates good fit between detector response and the sinusoidal function, as the fit indicates that $\theta_0=0.2^\circ\pm 0.3^\circ$ at $f=40.791$ GHz. This same method was used to extract $\theta_0$ at a number of  frequencies. These fit-extracted $\theta_0$ of the bare sample holder plus detector have been plotted as a function of $f$ in Fig.\ 1(d), which indicates that  $-8^\circ \le \theta_{0} \le 6^\circ$ for $32 \le f \le 50$GHz. Here, one expects that $\theta_{0} = 0$ if the microwave antenna is well aligned with the microwave detector when the $\theta=0$ condition is defined at each $f$. Thus, the observed spread in these $\theta_{0}$, which is $ \approx 14^\circ$, is attributed to experimental issues such as misalignment and readout errors.

\begin{figure}[t]
\includegraphics[width=75mm]{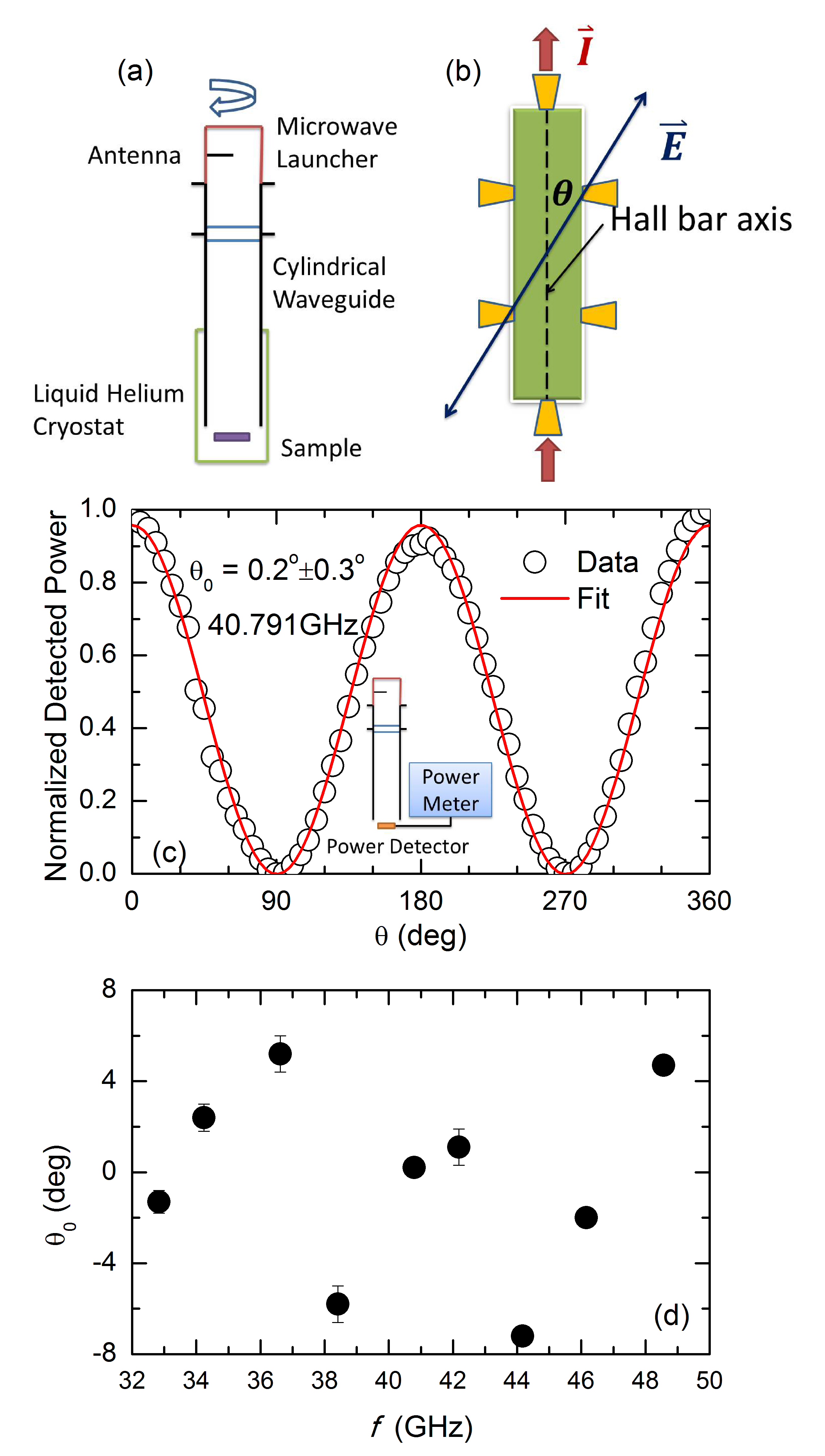}
\caption{(Color online) (a) Linearly polarized radiation generated by an antenna inside a rotatable microwave launcher is transmitted onto the specimen via a cylindrical waveguide. (b) This figure illustrates the orientation of the linearly polarized microwave radiation with respect to the specimen. $\theta$ is defined as the angle between the microwave electric field $E$ and the  Hall bar axis. (c) The sinusoidal normalized detected power as a function of $\theta$, with a power detector in place of the sample, can be fit by a cosine square function in order to extract $\theta_0$. (d) A plot of $\theta_0$ vs $f$, without the sample, implies that the measurement uncertainty in $\theta_0$ is below $\approx 14^\circ$.}
\end{figure}

\begin{figure}[t]
\includegraphics[width=75mm]{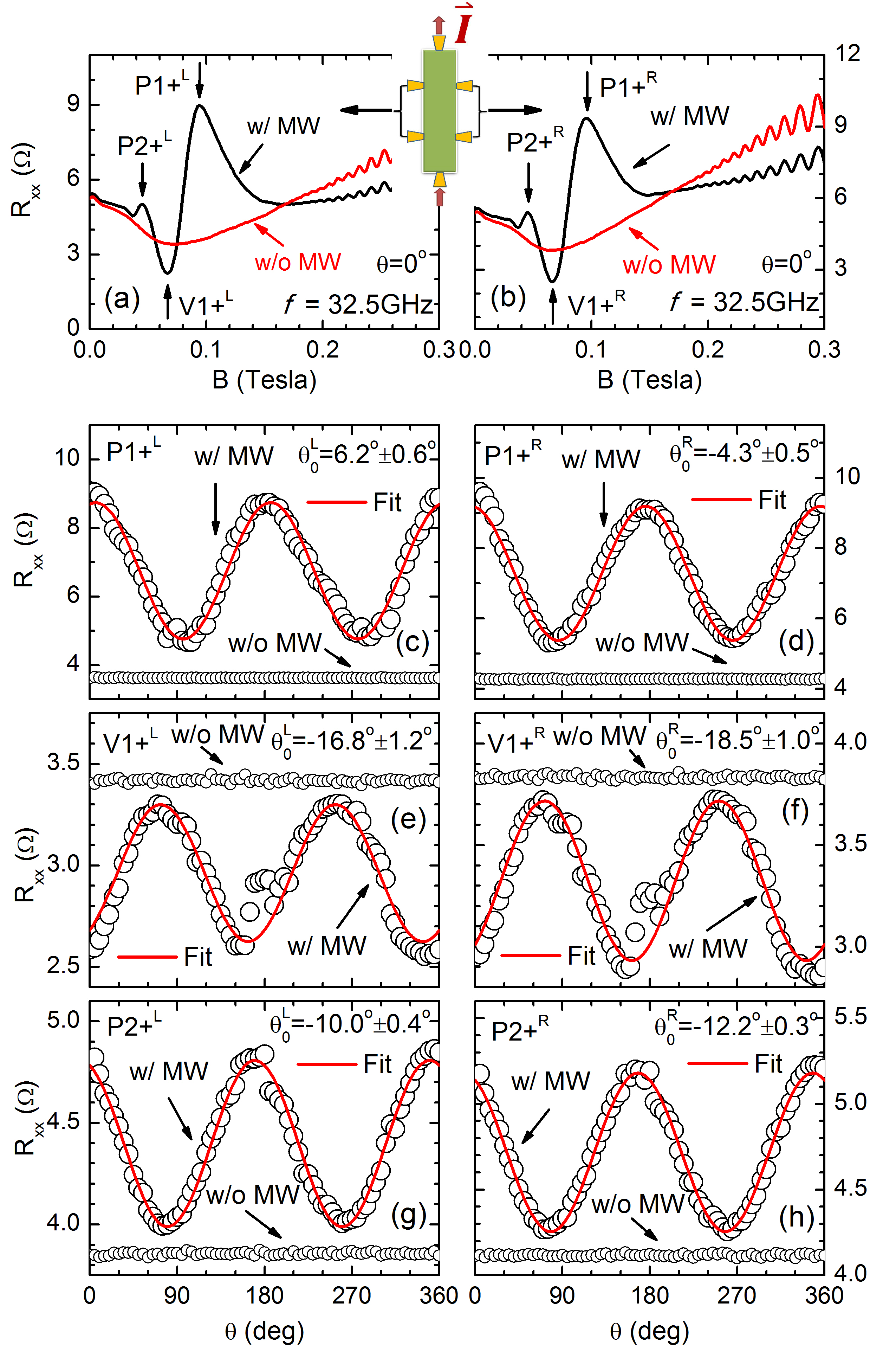}
\caption{(Color online) Panels (a) and (b) show $R_{xx}$ with (photo-excited) and without (dark) microwave illumination for $0 \le B \le  0.3$ T, at $f=32.5 GHz$ and $\theta=0$. Here, panel (a) shows the $R_{xx}$ measured from the left (L) pair of contacts while panel  (b)  shows the $R_{xx}$ measured from the right (R) pair of contacts on the Hall bar. Panels (c)-(h) exhibit oscillatory photo-excited and dark $R_{xx}$ vs. the polarization angle $\theta$, with extracted $\theta_0$ obtained from the  fit curve at (c) $P1+^L$ (d) $P1+^R$ (e) $V1+^L$ (f) $V1+^R$ (g) $P2+^L$ (h) $P2+^R$ respectively. Panels (c)-(h) show that the fit extracted $\theta_0$ at a given extrema of the MIMOs have similar values for contact pairs on opposite sides ((L) and (R)) of the device with the phase shift difference being smaller than estimated measurement uncertainty ($\approx 14^{0}$).}
\end{figure}

In the next experimental phase, $B$-field sweeps of $R_{xx}$ with microwave (photo-excited $R_{xx}$) and without microwave excitation (dark $R_{xx}$) were carried out from $-0.3$ T to $0.3$ T at a number of frequencies. Figs.\ 2(a), and  2(b) both show photo-excited and dark $R_{xx}$ vs $B$ with $f=32.5$ GHz, $P=0.63mW$ at $\theta=0^\circ$ for $0 \le B\le 0.3$T. Here, Fig.\ 2(a) (labeled with the superscript sign of L) represents measurements from the contact pair on the left side of the Hall bar, while Fig.\ 2 (b) (labeled with the superscript sign of R) represents measurements from the right side of the Hall bar.  Both Fig.\ 2(a) and 2(b) exhibit perceptible RIMOs below $0.15$ T. Thus, in Fig.\ 2(a) and (b), the predominant oscillatory extrema of RIMOs have been labeled as $P1+$, $V1+$, and $P2+$, to indicate the first peak, the first valley, and the second peak, respectively, of RIMOs for $B \ge 0$. Fig.\ 2(c)-(h) display dark and photo-excited $R_{xx}$ as a function of $\theta$ at fixed $B$ corresponding to $P1+^{L}$, $P1+^{ R}$, $V1+^{L}$, $V1+^{R}$, $P2+^{L}$, and $P2+^{R}$. Note that the photo-excited $R_{xx}$ varies sinusoidally with $\theta$, and the dark $R_{xx}$ maintains a constant value vs. $\theta$.

\begin{figure}[t]
\includegraphics[width=75mm]{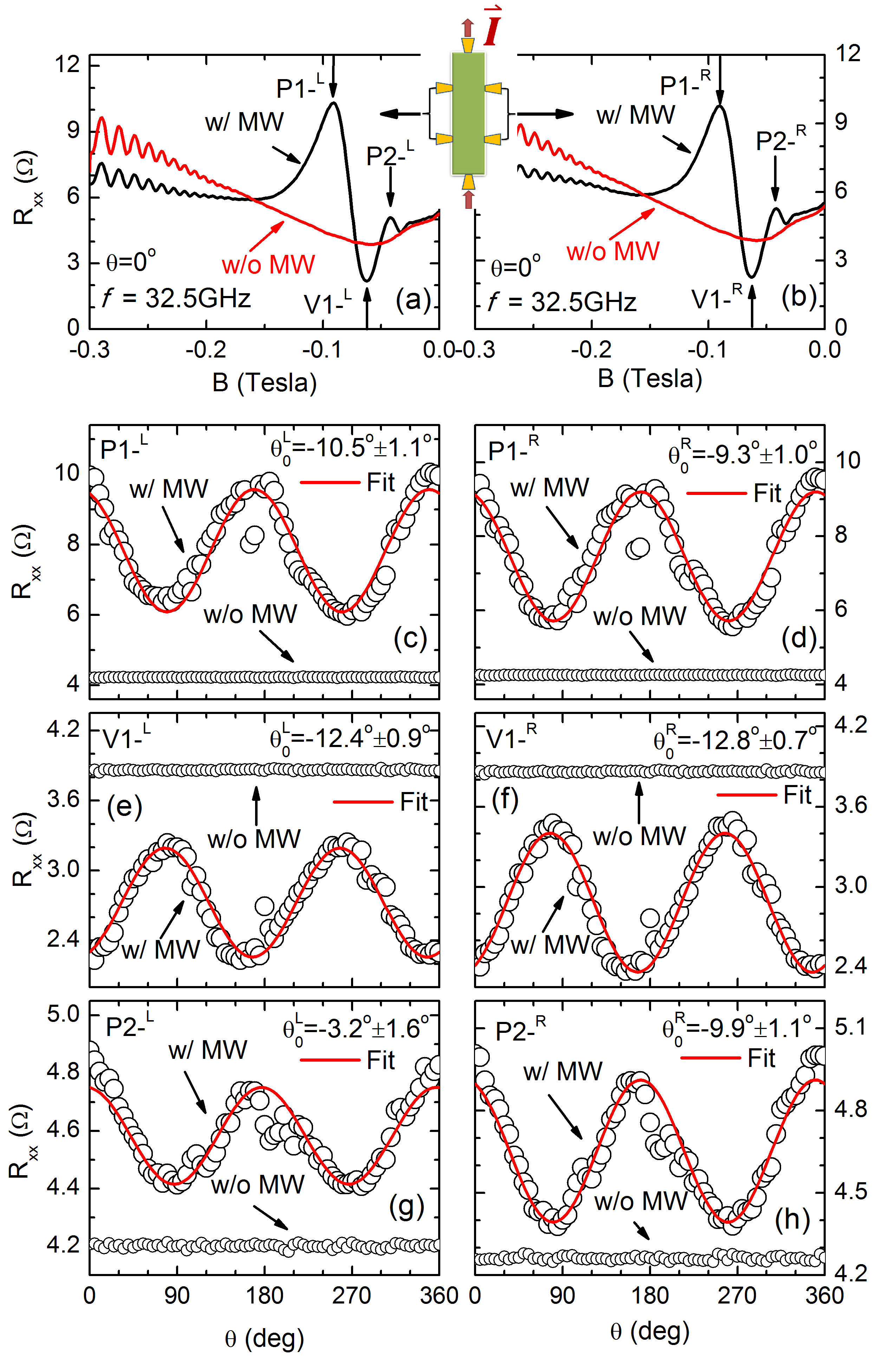}
\caption{(Color online)  Panels (a) and (b) show $R_{xx}$ with (photo-excited) and without (dark) microwave illumination for $0 \le B \le  0.3$ T, at $f=32.5 GHz$ and $\theta=0$. Here, panel (a) shows the $R_{xx}$ measured from the left (L) pair of contacts while panel  (b)  shows the $R_{xx}$ measured from the right (R) pair of contacts on the Hall bar. Panels  (c)-(h) show oscillatory photo-excited and dark $R_{xx}$ from the left-sided pair and the right-sided pair of contacts on the Hall bar,  along with the extracted $\theta_0$ at (c) $P1-^L$ (d) $P1-^R$ (e) $V1-^L$ (f) $V1-^R$ (g) $P2-^L$ (h) $P2-^R$, respectively. Panels (c)-(h) show that the fit extracted $\theta_0$ at a given extrema of the MIMOs have similar values for contact pairs on opposite sides ((L) and (R)) of the device with the phase shift difference being smaller than estimated measurement uncertainty ($\approx 14^{0}$).}
\end{figure}

\begin{figure}[t]
\includegraphics[width=75mm]{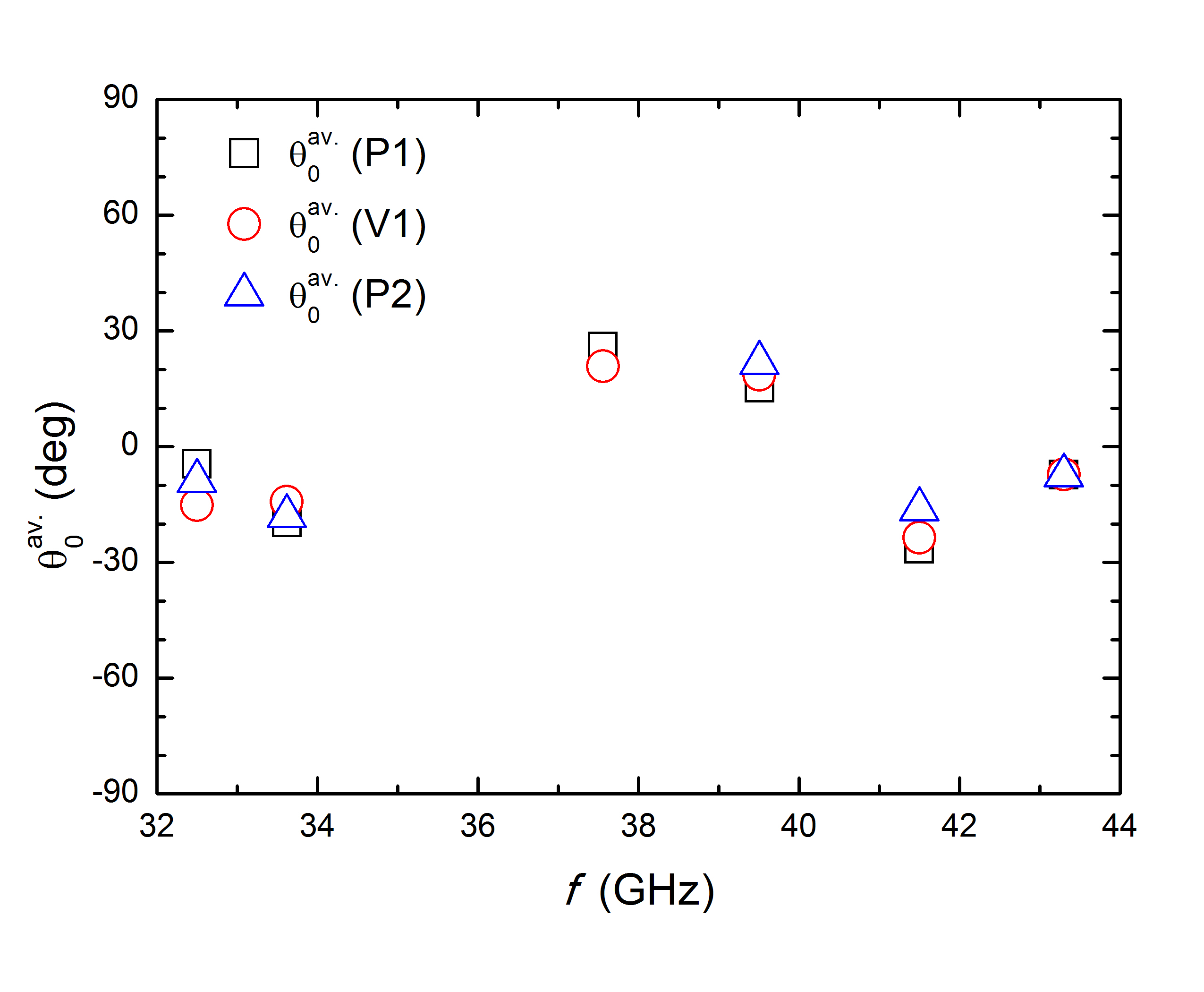}
\caption{(Color online)  This panel shows a plot of the average value of $\theta_0$, i.e., $\theta_{0}^{av.} =  (\theta_{0}^{+} + \theta_{0}^{-})/2$ vs $f$ for various extrema.  The figure shows that $\theta_{0}^{av.}$ for all extrema at a given $f$ are close to each other. However, there is a large variation in $\theta_{0}^{av.}$ with $f$.}
\end{figure}

The phase shifts $\theta_0$ in the $R_{xx}$ vs. $\theta$ data are extracted by fitting to $R_{xx}=A\pm C\cos^2(\theta-\theta_0)$, with  "+" sign for peaks and "-" sign for valleys. Associated fit curves are also shown in Fig.\ 2(c)-(h).  To compare these $\theta_0$, let us begin by focusing upon Fig.\ 2(c) ($P1+^L$) and Fig.\ 2(d) ($P1+^R$). Again, both of $P1+^L$ and $P1+^R$ are measured at the $P1+$ magnetic field, but from contact pairs on opposite sides of the Hall bar. The results show $\theta_0^{L}=6.2^\circ\pm 0.6^\circ$ for $P1+^L$ and $\theta_0^{R}=-4.3^\circ\pm 0.5^\circ$ for $P1+^R$, where the small standard errors are due to the excellent fits to the cosine square function for oscillatory $R_{xx}$.  The $\delta \theta_{0} =| \theta_{0}^{R} - \theta_{0}^{L}| = 10.5^\circ\pm 0.8^\circ$  between $P1+^L$ and $P1+^R$,  is less than the measurement uncertainty of $14^\circ$. Similarly, if we compare Fig.\ 2(e) ($V1+^L$) and (f) ($V1+^R$) or (g) ($P2+^L$) and (h) ($P2+^R$), the $\delta \theta_0 = 1.7^\circ\pm 1.6^\circ$ for $V1+$, and $\delta \theta_0 = 2.2^\circ\pm 0.5^\circ$ for $P2+$. These phase shift differences $\delta \theta_{0}$ are again smaller than the measurement uncertainty. 

Next, a comparison of the photo-excited and the dark $R_{xx}$ vs $B$ is shown in Fig.\ 3(a) and (b) for the same experimental parameters over the range $-0.3 \le B \le 0$ T. Similar to Fig.\ 2, $R_{xx}$ measured on the left contact pair on the Hall bar is plotted in Fig.\ 3(a) (with superscript L), and that measured via right contact pair is shown in Fig.\ 3(b) (with superscript R). Three prominent extrema, which are now labeled as $P1-$, $V1-$ and $P2-$, represent the first peak, the first valley, and the second peak, respectively, for $B \le 0$. Fig.\ 3(c)-(h) also exhibit traces of oscillatory photo-excited and dark $R_{xx}$ vs $\theta$ at the $P1-^{L}$, $P1-^{R}$, $V1-^{L}$, $V1-^{R}$, $P2-^{L}$, and $P2-^{R}$ magnetic field values indicated in Fig.\ 3(a) and (b). The $\theta_0$ extracted by fitting the data to $R_{xx}=A\pm C\cos^2(\theta-\theta_0)$ are also displayed in Fig.\ 3 (c)-(h). As for the data of Fig.\ 2, we compare the $\theta_0$ of opposite contact pairs at the extrema of RIMOs in the range of $B \le 0$. At $P1-^L$, $\theta_0^{L}=-10.5^\circ\pm 1.1^\circ$, and at $P1-^R$, $\theta_0^{R}=-9.3^\circ\pm 1.0^\circ$, which yields $\delta \theta_{0} = |\theta_{0}^{R} - \theta_{0}^{L}| = 1.2^\circ\pm 1.5^\circ$, still much smaller than the measurement uncertainty. For $V1-$ and $P2-$ (Fig.\ 3(e)-(h)), the $\delta \theta_0 = 0.4^\circ\pm 1.1^\circ$  and $\delta \theta_0 = 6.7^\circ\pm 1.9^\circ$, respectively, which are also within the estimated measurement uncertainty of $\approx 14^\circ$.    

Similar results to those shown Fig.\ 2 and Fig.\ 3 for $f=32.5$ GHz were observed at other microwave frequencies. That is, each $\delta \theta_0$ at these $f$  were smaller than the measurement  uncertainty.  As a consequence, we make the assumption that the $\theta_0$ values obtained from the two opposite contact pairs at given $B$ in this sample are practically indistinguishable. Thus, we average the $\theta_0^{R}$ and $\theta_{0}^{L}$ obtained from the opposite contact pairs of the Hall bar to reduce the measurement error in the extracted $\theta_0$, and assume that this average $\theta_0$ is more representative of the corresponding sample area  (see Fig.\ 1(b)). Table 1 provides a summary of the $\theta_{0} = (\theta_{0}^{R} + \theta_{0}^{L})/2$ obtained after averaging over opposite contact pairs. Here, $\theta_{0}^+$ and $\theta_{0}^-$ are the $\theta_{0}$'s for positive and negative magnetic fields, respectively.

\begin{table*}[t]
\caption{The representative $\theta_0^{+}$ and $\theta_{0}^{-}$ at various $f$, at the extrema of RIMOs. Here, the superscripts '+' and '-' refer to positive and negative magnetic fields. $\theta_0^{+}$ and $\theta_{0}^{-}$  at $f=37.56$ for $P2$  are missing because these peaks were too small to be measured reliably. This comparison implies a given extremum under field reversal, shows similar $\theta_0$ values for each $f$.}
\begin{ruledtabular}
\begin{tabular}{crrrrrr}
$f$ (GHz)&$\theta_0^+$ $(P1)$&$\theta_0^-$ $(P1)$&$\theta_0^+$ $(V1)$&$\theta_0^-$ $(V1)$&$\theta_0^+$ $(P2)$&$\theta_0^-$ $(P2)$\\
\hline
32.50&$1.0^\circ\pm 0.4^\circ$&$-9.9^\circ\pm 0.7^\circ$&$-17.7^\circ\pm 0.8^\circ$&$-12.6^\circ\pm 0.6^\circ$&$-11.1^\circ\pm 0.3^\circ$&$-6.6^\circ\pm 1.0^\circ$\\
33.62&$-12.5^\circ\pm 0.6^\circ$&$-26.8^\circ\pm 0.6^\circ$&$-11.2^\circ\pm 0.5^\circ$&$-17.5^\circ\pm 0.9^\circ$&$-14.0^\circ\pm 0.4^\circ$&$-22.0^\circ\pm 0.8^\circ$\\
37.56&$25.1^\circ\pm 0.4^\circ$&$26.7^\circ\pm 0.5^\circ$&$25.0^\circ\pm 0.5^\circ$&$16.7^\circ\pm 0.5^\circ$&\multicolumn{1}{c}{$-$}&\multicolumn{1}{c}{$-$}\\
39.51&$16.3^\circ\pm 0.6^\circ$&$14.3^\circ\pm 0.8^\circ$&$11.7^\circ\pm 1.1^\circ$&$25.7^\circ\pm 0.9^\circ$&$18.4^\circ\pm 0.7^\circ$&$25.1^\circ\pm 0.8^\circ$\\
41.50&$-32.5^\circ\pm 0.6^\circ$&$-20.5^\circ\pm 0.4^\circ$&$-23.3^\circ\pm 1.0^\circ$&$-23.8^\circ\pm 1.0^\circ$&$-18.5^\circ\pm 1.8^\circ$&$-14.0^\circ\pm 0.6^\circ$\\
43.30&$-11.7^\circ\pm 0.3^\circ$&$-2.7^\circ\pm 0.5^\circ$&$-11.1^\circ\pm 0.8^\circ$&$-3.2^\circ\pm 1.9^\circ$&$-14.3^\circ\pm 0.9^\circ$&$-0.6^\circ\pm 0.9^\circ$\\
\end{tabular}
\end{ruledtabular}
\end{table*}

For the sake of extracting the frequency dependence in $\theta_{0}$, we evaluate also $\theta_{0}^{av.}$, which is the average of  $\theta_0^+$ and $\theta_0^-$. For example, $\theta_{0}^{av.}(P1)= (\theta_0^+(P1) + \theta_0^-(P1))/2$, etc. The $\theta_0^{av.}$ at the various extrema are shown in Fig.\ 4 as a function of $f$. Fig.\ 4 shows that  the differences between any two of  $\theta_0^{av.}$ are less than $14^\circ$ at each $f$. Further, the points appeared clustered at each $f$, and the point-clusters show a systematic variation with $f$. These results indicate a frequency dependence in the phase shift $\theta_{0}$ even when a worst case scenario is applied for averaging over other smaller contributions to the phase shift.

\section{discussion} 
From the results presented above, we suggest the following: (a) At low $P$, opposing contact pairs on a Hall bar present the same value for a given extremum, a given magnetic field direction, and a given $f$, for $\theta_0$ that is extracted from the fit of $R_{xx}(\theta)=A\pm Ccos^2(\theta -\theta_0)$, within measurement uncertainty.  (b) At a given $f$, the $\theta_{0}^{av.}$ for all the considered extrema exhibit similar values, see Fig.\ 4. However, these values show a systematic dependence upon $f$.  

Note that point (a) is expected since the two edges of the Hall bar are parallel to each other and their orientation with respect to the microwave polarization is the same. Since the GaAs/AlGaAs heterojunctions represent an extraordinarily clean system with mean-free paths approaching the mm- or sample size-scale, it is difficult to develop a scenario where the two edges of the homogeneous specimen would not exhibit the same linear polarization response. In this context it might be interesting to introduce non-parallel edges to see if that feature in the specimen introduces a phase shift in the response observed on the two edges.  This will be a topic of future experiments.

Point (b) suggests that  the magnitude of $B$ does not produce profoundly distinguishable differences on representative $\theta_0$ at the extrema of RIMOs in the specimen. The difference in this aspect between present and previous experimental work could be attributed to differences in sample quality and defect configuration within the specimen.  From the theoretical perspective, Lei~\textit{et~al.}\cite{2012XLLeiPRB}~ have simulated sinusoidal responses of $R_{xx}$ as a function of polarization angle $\theta$ using the balance-equation formulation of their photon-assisted magnetotransport model. The results have indicated that the phase shift in the $R_{xx}$ vs. $\theta$ response is dependent upon $B$ or extremum, and $f$. Further, they suggested that $P1+(\theta)=P1-(\pi-\theta)$, $V1+(\theta)=V1-(\pi-\theta)$, etc.~in an isotropic system, which is not observed here. However, it should be noted that real samples possibly include additional complexity, such as asymmetry, that was not considered in their theory.\cite{2012XLLeiPRB}
 
Finally, although the cause of observed frequency-dependent $\theta_0$ is not yet fully understood experimentally, and theory has  not considered this possibility in our context,  we draw a comparison with Faraday rotation in quantum Hall systems. In Faraday rotation,  the polarization-plane of the transmitted linearly polarized radiation in the presence of magnetic field becomes a rotated by an angle, $\theta_F$, which is called the Faraday angle. Generally, Faraday rotation, $\theta_F$, is a function of radiation frequency per the Drude-Lorentz model\cite{2011ICrasseeNaturePhysics,2011AMShuvaevPRL,2013RShimanoNatCommun}.  The  phase shifts $\theta_{0}$ reported here appear, however, in a $dc$ response, the magnetoresistance, observed under $ac$-excitation. One might qualitatively understand the observed $\theta_{0}$ by suggesting that the 2DES rotates the polarization of the $ac$ excitation, and the $dc$ response then follows polarization of the rotated $ac$ excitation. In such a situation, $\theta_{0}$ could be a manifestation of $\theta_{f}$. Further theory is needed, however, to provide more understanding of this possibility.

\section{conclusion} 
The phase shifts observed in the polarization-angle-dependence of microwave-induced magnetoresistance oscillations have been processed using an analysis  which emphasized averaging over the smaller contributions, when the smaller contributions were smaller than estimates of the experimental uncertainty. The analysis was carried out in order to extract a possible  microwave frequency contribution to the phase shift, $\theta_{0} (f)$, observed in the $R_{xx}$ vs. $\theta$ response of the microwave radiation-induced magnetoresistance oscillations. The analysis demonstrates a non-vanishing frequency dependent contribution to the phase shift over the frequency interval $32 \le f \le 50$ GHz. 
\section{acknowledge}
Magnetotransport studies and T.Ye at Georgia State University were supported by the US Department of Energy, Office of Basic Energy Sciences, Material Sciences and Engineering Division under DE-SC001762. H.-C. Liu was supported by the ARO under W911NF-10-1-0450.

\end{document}